# COMPARATIVE REVIEW OF MALWARE ANALYSIS METHODOLOGIES

Ioannis G. Kiachidis and Dr. Dimitrios A. Baltatzis

School of Science and Technology, International Hellenic University, Thermi - Thessaloniki, Greece

## ABSTRACT

*To fight against the evolution of malware and its development, the specific methodologies that are applied by the malware analysts are crucial. Yet, this is something often overlooked in the relevant bibliography or in the formal and informal training of the relevant professionals. There are only two generic and all-encompassing structured methodologies for Malware Analysis (MA) – SAMA and MARE. The question is whether they are adequate and there is no need for another one or whether there is no such need at all. This paper will try to answer the above and it will contribute in the following ways: it will present, compare and dissect those two malware analysis methodologies, it will present their capacity for analysing modern malware by applying them on a random modern specimen and finally, it will conclude on whether there is a procedural optimization for malware analysis over the evolution of these two methodologies.*

## KEYWORDS

*Malware Analysis, SAMA, MARE, Methodology, Review, Comparison.*

## 1. INTRODUCTION

Cybercrime is considered one of the most detrimental menaces of the modern world, especially when it comes to the finance sector. There is a projection of a rising cost of cybercrime to around 10.5 trillion USD by 2025 [1]. Integral part of the above "success" is malware. During 2020 there was approximately one ransomware case every 11 seconds against a business worldwide [2]. The outbreak of the COVID-19 pandemic led to the increase of attacks, mainly phishing attempts, by 600% [3]. In the last 10 years malware growth was increased by almost 70 times and amounts to more than a billion infections [4]. With insurmountable numbers of new variants and novel threats being released worldwide daily, it is at least challenging and usually impossible for the analysts to catch up.

Beyond all, analysis needs to be precise, trying to achieve the highest level of accuracy possible, and as less time consuming as possible. As one more of the countless problems faced, it needs a methodology developed to tackle it. There were no structured methodologies developed until 2010 when the MARE methodology (Malware Analysis Reverse Engineering) [5] was presented. This can be considered the only available methodology and go to solution until recent days. Still, it was developed in a different time with different threats roaming the landscape. While trustworthy there was a need of something contemporary to tackle the challenges presented in the modern battlefield. In 2020, the SAMA methodology (Systematic Approach to Malware Analysis) [6] was presented. SAMA was created by the aforementioned need for a new procedural model designed to face the newly emerged threats. As a methodology, it tries to remedy the weaknesses of the MARE while preserving its strong points. In addition, it was designed and adapted as a modern analysis procedure and is battle-tested against formidable malware by its creators to provide the needed Proof-of-Concept.

Yet, a methodology is not enough on its own. Methodology provides the steps needed to proceed when following a procedure with a specific goal set beforehand. The specific tools and capabilities for each step and in between, are those that can make a difference both in terms of accuracy and speed.

Beyond the established and well-known parts of MA (i.e. static and dynamic analysis techniques) the modern palette contains some interesting, innovative and capable solutions regarding MA but which are not generic and were made to cover specific needs:

- Automated Analysis (sandboxed VMs used for automated dynamic analysis for samples) [16][34]
- Hybrid Analysis (combination of static and dynamic analysis) [8][23][25]
- Memory analysis (acquisition and examination of the system's memory dump) [9][10][11]
- Side-Channel analysis (a novel and dynamic analysis technique used primarily on IoT) [12][13][14][15][18]
- Machine Learning (the utilization of ML techniques in MA) [17][19][22][24]
- Deep Learning (the incorporation of DL techniques in MA) [19]

## 2. BACKGROUND WORK

Regarding malware analysis and its components there is quite a number of research work epitomized in MA.

On Practical Malware Analysis [20], Sikorski M. and Honig A. present the process of malware analysis in a detailed and thorough manner. Both static and dynamic MA are analyzed and presented including the process of code analysis.

In a similar fashion, Monnappa K. A. in Learning Malware Analysis [7] presents a more modern equivalent of [20]. Static and dynamic malware analysis are still analyzed here thoroughly. The main difference is the fact that this book uses more modern examples of systems and malware samples.

There is also some research revolving around literature review which provides access to crucial definitions on malware analysis during years past. One example is Verma A. et al, A Literature Review on Malware Analysis [21]. This is a brief anthology of modern research on MA and can be used as an indicator of the evolution of the aforementioned research.

A more detailed work on MA (like dynamic analysis and parts of it) is the Dynamic Malware Analysis in the modern Era – A State of the Art Survey by Or-Meir O. et al [12]. This paper is a great review of the modern landscape regarding dynamic analysis and it additionally provides a brief overview of ML methods used for dynamic MA.

Regarding Machine Learning in malware analysis specifically, Gibert D. et al. presented the - "The rise of machine learning for detection and classification of malware: Research developments, trends and challenges" [19]. It provides a review regarding ML techniques utilized in MA while presenting the drawbacks of the existing ML methods and analyzing the current and future development of ML in MA.

In addition to the above, the paper MMALE — A Methodology for Malware Analysis in Linux Environments by José Javier de Vicente Mohino et al. [33] presents a novel methodology of MA for Linux environments under the prism of IoT. MMALE methodology is an amalgamation of existing MA methodologies and it utilizes, mainly, parts of SAMA methodology. MMALE methodology can act as a proof of the dynamic and potential of the SAMA methodology.

This paper is revolving around the two generic methodologies of MA, SAMA and MARE. SAMA can be considered as an evolution of MARE. Thus, the most recent model will be presented and applied and this is the part that is associated with the SAMA methodology and its subsequent comparison with MARE. On this specific subject there is no research available at the time of writing and this is our intention.

## 3. PRESENTATION OF METHODOLOGIES

### 3.1. MARE (Malware Analysis Reverse Engineering)

Until the dawn of the second decade of the $21^{st}$ century, there weren't any structured methodologies available for MA. MARE [5] was created to fill that gap. The researchers of MARE developed a Malware Defense (M.D.) timeline which outlines the goals of the research and which is part of the original work. The general outline of M.D., after the infection occurs, is Detection, Isolation and Extraction, Behavior Analysis, Code Analysis and Reverse Engineering, Pattern Recognition, Malware Inoculation and Remedy.

MARE resides between the Detection and Code Analysis and Reverse Engineering phases of M.D. As the researchers themselves stated, MARE introduces logical steps taken in each process to help analysts produce an output that is repeatable, objective and applicable, all of the aforementioned with the purpose of a better understanding of the analyzed malware.

Detection phase is the first phase encountered after the infection according to M.D. During this phase the malware in question is scanned by malware scanners e.g., VirusTotal. This is done for verifying whether or not the malware presents a threat already encountered or an unknown one (0-day). The verification is done based on the signature derived by the cryptographic hash of the malware. This process may not present trustworthy results when dealing with highly sophisticated and advanced threats. Also, there is a chance that the results will flag the malware as non-malicious and this can act as a trigger for further researching the root of the disruption that made the whole process start in the first place.

Isolation & Extraction phase follows the detection phase. This phase's objective is to isolate, extract and secure the malware. This is because of the need to securely transfer the malware to an environment for the behavioral analysis. The malware is located in the infected system and it is then extracted. After the extraction, it is compressed, and a password is set. It is important for the compressed malware to be password protected because there is always the possibility that the malware will auto-execute. In addition, during this step, analysts try to gain insight of the malware's nature and try to figure out whether the malware is a rootkit or not. In the case of a rootkit, the extraction method differs.

Next comes the Behavioral Analysis phase. During this phase analysts try to observe and take note of the changes that the execution and functionality of the malware makes to the system. This is done because changes made by the malware are the hint towards figuring out its malicious purpose. These changes that analysts must consider, are file manipulation, registry tampering, library modification, connections initiated etc. Analysts take snapshots prior to the execution of the malware and after its execution. The differences between these snapshots are indicative of the changes made due to malware functionality. Automated dynamic analysis is also useful for this step. It helps speeding up the whole process and based on the results analysts may further manually examine the malware (e.g., when the malware seeks for user interaction). This step may need to be repeated, based on findings derived from the next step of Code Analysis & Reverse Engineering.

Last comes the Code Analysis & Reverse Engineering phase. This phase is heavily dependent on assembly language utilization. It revolves around debugging and disassembling. The core focus is the understanding of the inner workings of the malware from the code point of view. Usually, it starts with the identification of strings and continues from there on. This step reveals a lot when done by competent analysts. Thus, malware developers deploy countermeasures to assure that analysts will have a difficult time during this step, to the point that this step is even nullified. Methods of anti-debugging, encryption and obfuscation are the tools in the hands of malware developers that help them protect their product from being reverse engineered. During this step analysts may find hints that can potentially lead them back in executing again the behavioral analysis step.

### 3.2. SAMA (Systematic Approach to Malware Analysis)

Almost 10 years after MARE was published, a second structured methodology for MA was developed. The newly created methodology was named SAMA [6]. Given the fact that attack vectors have increased immensely during those years it seems that the MARE methodology could be deemed as inadequate for the present needs. While complete in terms of presenting the required phases to thoroughly carry out the MA procedure, MARE lacks the capacity for addressing challenges that emerged from the expansion of the deployed techniques complexity. MARE comprises of 4 phases, namely Detection, Isolation and Extraction, Behavioral Analysis, Code Analysis and Reverse Engineering as already stated.

The SAMA approach tries to standardize the MA procedure to cope with the needs of the contemporary challenges. This new approach retains the 4 phases which are now named as follows:

1. Initial Actions
2. Classification
3. Static and Dynamic Code Analysis
4. Behavioral Analysis

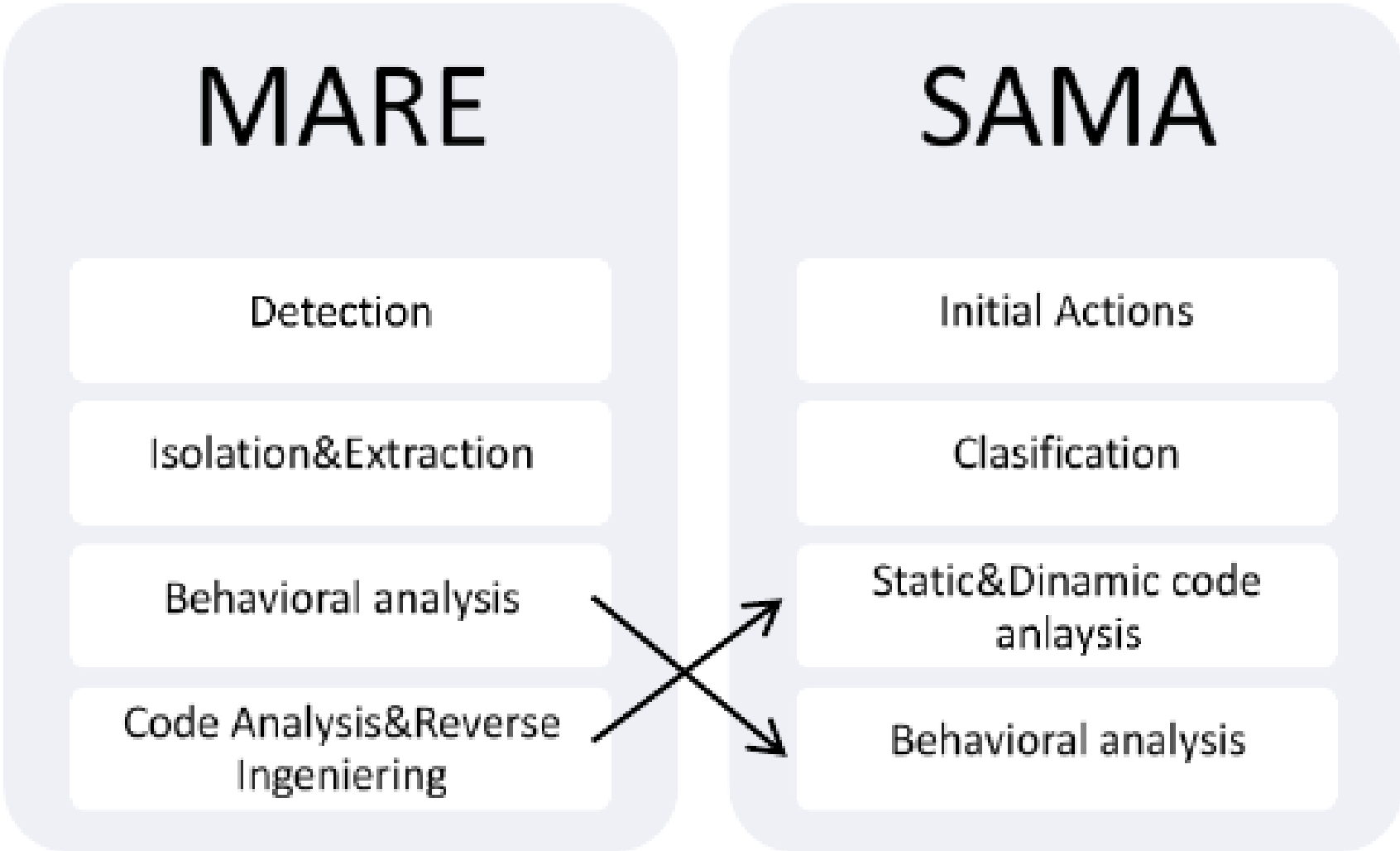

Figure 1. MARE and SAMA phases comparison [6]

The aim of the new procedure is the provision of a framework capable of analyzing modern and complex malware. The demand for such a framework is the need for an iterative process which is systematic, structured, founded upon solid and rigid methods, able to assist the analysts in acquiring the knowledge necessary from a specific malware. This knowledge is derived from the information gathered throughout the analysis and is crucial for understanding the operations of a sample and as a result the possible methods of containing and removing it. Through the completion of the analysis procedure by following the four main phases, SAMA tries to figure out and carve all the needed information to finally defeat the malicious adversary.

In the Initial Actions phase of SAMA, analysts should write down the state and the form of the systems that will be used to carry out the analysis procedure. Systems can be either virtual or physical. In either case snapshots must be taken so as the possibility to revert to a clean state is reserved by the analyst. Also, the integrity of the aforementioned snapshots must be secured and thus the relevant hash values must be generated and retained. Until now the analyst has yet to begin the analysis of the actual sample in question. It can be argued that initial actions are just a preliminary phase, focused in security of the analysis environment but this phase is mandatory and a prerequisite for the analysis procedure itself.

Next comes the Classification phase. For the completion of this phase, basic static analysis steps, form an adequate procedure, which is a prerequisite for the phases to follow. The code of the sample is not examined in this phase and the main scope is the confirmation of whether there must be a continuation of the analysis process or not because until now, there are no concrete indications that the sample is either benign or malicious. The tasks of this phase include the transfer of the sample to the analysis environment, hashing of it for reasons of identification (there is the possibility for the sample to be a known threat), the collection of information from OSINT sources regarding the nature of the sample (e.g. the definition of a possible malicious sample's family by the use of the VirusTotal online engine), the collection and analysis of the sample strings, the definition of the possibility for obfuscation techniques applied to it (e.g. by the use of entropy measurement or the detection of packaging deployed to it by using tools like PEiD) and the analysis of the sample's format.

The Code Analysis phase follows next, and in this phase advanced static and dynamic analysis techniques are deployed for achieving the goal of analyzing the code of the sample. This is most likely the most difficult and complex process for the analysts to complete. Unfortunately, for the modern malware which are highly sophisticated and complex, the completion of this phase is a necessity because its findings provide a great insight of how the sample operates and brings into light hidden paths of execution or obscure features that otherwise remain unseen. The techniques deployed are disassembly and debugging. Because debugging is a dynamic process, after the code analysis phase concludes, the analysis environment must be reverted back to its original state to continue towards the next phase.

The conclusion of SAMA comes with the Behavioral Analysis phase. As the name implies, in this phase dynamic and memory analysis techniques are deployed. The sample is executed in a safe analysis environment and all the changes caused by it (registry changes, connections established etc.) on the system are recorded and analyzed. Dumps of the memory of the system are gathered and the memory analysis commences to discover even the last glimpses of information that are left to be found.

Closing this part, it is of utmost importance to note that methodologies like the ones mentioned above are not locked into using only a specific arsenal of tools. This is a prerequisite because the availability of specific tools cannot be guaranteed in the long run and this fact must be always taken into consideration. A variety of tools must be compatible to a methodology and that is something that the designers of a methodology must ensure. Agility is better than rigidness in regard to this part.

## 5. METHODOLOGIES COMPARISON

Following the flow of the procedural execution of the SAMA mode, we are going to compare it against the MARE process. It is wise to do so since their differences can be made more apparent and thus useful conclusions may arise.

As seen already, SAMA starts with the Initial Actions phase. In this phase analysts, mainly, build and acquire the base configuration before the actual analysis of the malware sample commences. This is crucial because it provides the means to compare the outcome of the malware actions after its execution is completed, giving an established reference point during the Initial Actions phase. This phase is not present in MARE. MARE starts with the Detection phase. At this point there is an optimization favoring SAMA over MARE because detection is a process that precedes the MA chronologically (i.e. it is essentially part of the forensics procedure prior to MA procedure). Some may argue that these are also distinct parts of another, more thorough procedure (i.e. the whole process begins from realizing that there is an infection and indicates MA application over the sample). The fact is that they are not the same in essence and that detection has enormous costs associated with it e.g., hardware and software infrastructure for IDPS, AVs, firewalls etc. Also, there is a high probability that the malware would be one obtained from third parties (i.e. there is no detection taking place). Ultimately, that makes detection irrelevant to MA.

Next to come in SAMA is the Classification phase. The SAMA model defines this part as the part related to the actual static analysis of the sample in question without the inclusion of the advanced static analysis (disassembly). MARE on the other hand, continues with the Isolation & Extraction phase and throughout this phase this goal is to locate, extract and transfer of the sample. Once again, this phase loses its relevance, although not entirely, when analysts are to analyze samples already handed over by third parties. The process is not entirely irrelevant because it is important to avoid the automatic execution of the sample. Still, it seems that this phase is actual a small part of the modern approaches to MA and it is not efficient to be considered a complete process but rather a small part of existing ones**.** It must be also noted that during the period MARE was developed, the most fearsome adversary were rootkits and researchers tried to develop a methodology that revolved around the dominant threat at the time (i.e. the design of the Isolation & Extraction phase of MARE is optimized for the containment and transfer of rootkits prior to the MA). While still powerful, nowadays threats like APTs have upped the ante to the game and thus this case makes the declared Isolation & Extraction phase obsolete (for the most part of it as already explained).

Code Analysis phase follows in SAMA. In this phase analysts commence the advanced static and dynamic analysis procedures (i.e., disassembly and debugging procedures). The main argument for this choice, which seemed quite odd since there was no basic dynamic analysis already completed, was the fact that the dynamic analysis procedure can actually be enhanced if it comes after the code analysis. That is because some parts of the sample's functionality may be revealed only through code analysis and this can be later used as an asset in dynamic (behavioral) analysis. MARE continues with the Behavior Analysis process. This is actually what its name implies. It is a basic dynamic analysis applied to the sample. While still quite relevant and useful, MARE presents its disadvantage to the very fact that SAMA researchers chose to place Code Analysis before Behavioral Analysis, something that has already been explained.

To conclude, SAMA ends with the Behavioral Analysis step. This is where the dynamic analysis takes place and the sample's behavior is observed. Conclusion scan be drawn based on the changes that took place during the execution of the malware. It is important to note that SAMA, since it was developed recently, it makes an allusion of memory analysis. This is something that is necessary because memory analysis, despite its drawbacks, may be the only analysis option available to the analysts when it comes to specific, contemporary, sophisticated foes. MARE on the other hand concludes with the Code Analysis and Reverse Engineering phase. This phase is, once again, what its name suggests and that is the application of advanced static and dynamic analysis techniques to the target (disassembly and debugging). It is easy to understand that SAMA and MARE become more or less the same towards their end with the main difference that their last two processes are reversed. This is another optimization favoring SAMA. Findings that arise throughout Code Analysis may give useful hints towards exploiting capabilities or weaknesses of the malware (e.g. the may be specific values that when used in the environment that the malware is designed to operate, may trigger some exceptional functionality of it and thus making it easier to understand the full spectrum of its nature). Also, there is no reference to memory analysis in MARE and this can be attributed to the relevant trends at the time that it was developed.

## 5.1. Comparison from a practical point of view

Onwards, a practical comparison of the late SAMA methodology will be done against MARE. A random Malware will be analyzed according to SAMA and at the same time a comparison will take as to what would have happened if the MARE methodology chosen.

A recent and contemporary malware sample was chosen for the procedure. It was randomly selected from a free and open database of malware samples found in dasmalwerk.eu site [26]. There was no prior knowledge of the sample itself or its nature. This was done to ensure the validity and integrity of the methodologies yielded results as a function of the fact that the sample was a mint one for those that carried out the analysis procedure. The same methodology steps would have taken place regardless of the sample in question and thus they can be repeated effectively if another specimen is chosen. The sample that was chosen is Gen:Variant.Johnnie.97338 (first seen on 2018, Johnnie variants are roaming wild in vast numbers still). No notion or idea behind its inner workings is available (not at least in a free and open form). The SHA-256 hash value of the sample is 240387329dee4f03f98a89a2feff9bf30dcba61fcf614cdac24129da54442762. Information about

the sample itself can be found in relevant VirusTotal and relevant sites using this hash value. This is a Malware designed for attacking Windows OS systems.

The analysis took place on a Linux host system [27], utilizing a Windows 10 guest VM [28]. A vast variety of tools was used to carry out the analysis. Examples are PEStudio, FakeNet-NG, PEview, PEid, IDA Free, x32/x64dbg, DIE, Process Explorer, RegShot.

### 5.1.1. Initial Actions

Throughout this phase we prepared the ground for the analysis. The relevant chain of actions was the following:

- Updated both host and guest systems. Both in terms of OS and tools that are to be used for the procedure.
- Took a clean snapshot of the guest (victim) system so as to reserve the capability of reverting back to a clean state.
- For the purpose of this paper, we acquired the sample specimen for the analysis from an open and freely available source (dasmalwerk.eu). As a result, for reasons of integrity and of course for identification and verification we hashed the sample and compared the values against those presented in dasmalwerk.eu.

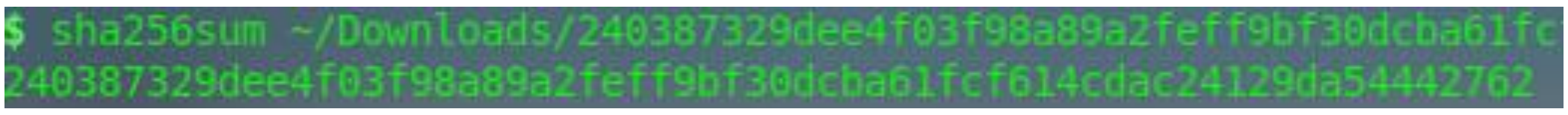

Figure 2. Finding sample's SHA-256 hashusing Linux terminal

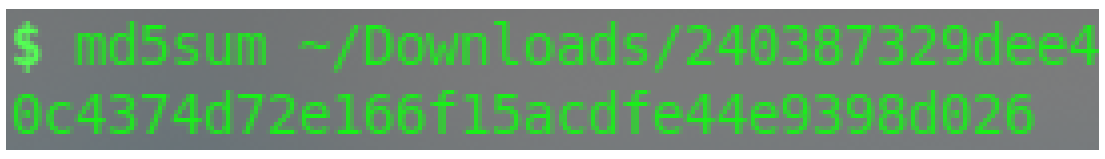

Figure 3. Finding sample's MD5 hashusing Linux terminal

- Isolated the victim system to ensure that there is no way malware spreading to other systems/subsystems (e.g., disabling shared clipboard features and setting a host-only adapter for the guest system to use).
- The victim system was booted and a system baseline snapshot was taken for having a point of reference to compare in post-analysis phases.

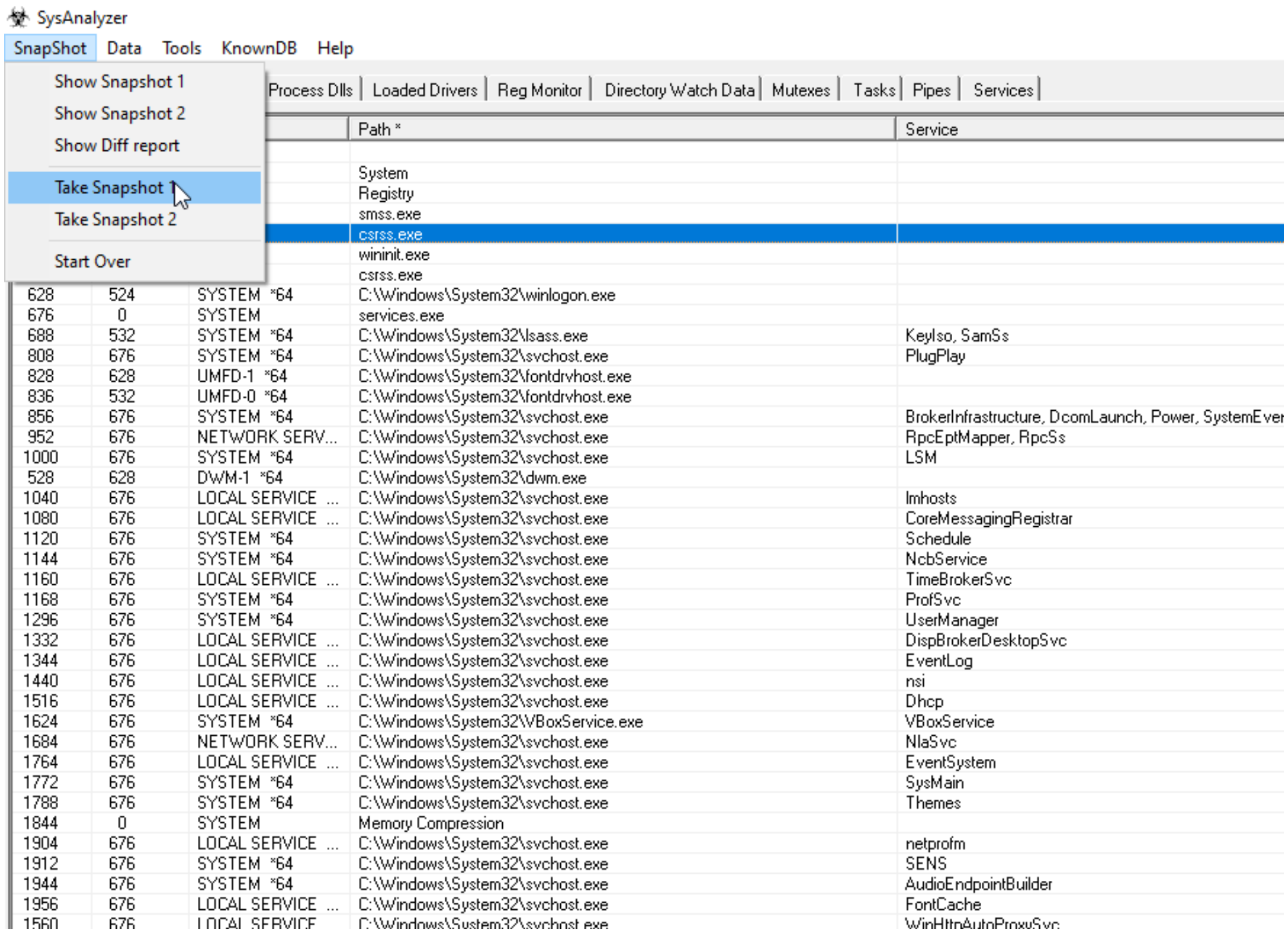

Figure 4. System baseline snapshot using SysAnalyzer

In comparison, MARE begins with the Detection phase. Detection is the second stage of M.D. (Malware Defense) Timeline. It seems that detection itself is closely related to stages preceding analysis and is cut from the analysis procedure. That being said, it is easy to conclude that MARE, at the time that it was developed, tried to be more generic i.e., it did not define the standards of the MA procedure specifically but it tried to be more involved in the M.D. Timeline as a whole. This means that SAMA gains a point here.

### 4.1.2. Classification

Classification phase starts with the secure transfer of the specimen in the analysis system which is isolated. It is important to remember that by the time the transfer is complete analysts must close any holes created for the transfer to take place. As a result, the analysis system will return to its isolated state which is really important for security reasons.

- Once the malware was inside the analysis system and the system was once again isolated, the malware sample was hashed again for reasons of integrity verification. This ensured that nothing happened to it during the transfer.
- Next, we checked whether the threat is a known threat or it is a zero-day. We knew of course that it is a known threat (after all we downloaded it from dasmalwerk.eu) but this is a presentation of the SAMA procedure and one of the steps it dictates is the aforementioned one. For this check we used VirusTotal free tier of services [29]. Not surprisingly, we were returned with the relevant results of the already known threat.
- Then we started the dissection of the specimen. To find any available information on the "surface" of it we used PEStudio for finding useful information about the nature of the threat. Specifically:
    - It is a Windows Portable Executable (4D 5A – MZ).

- It has an entropy of 3.4 which indicates a normal, unpacked and unencrypted file. This assumption is further enhanced when examining the details regarding the executable sections. Each raw and virtual size of each section is almost the same.
- It is developed for 32-bit architecture.
- It is a GUI executable.
- It was compiled on 5/9/2018.

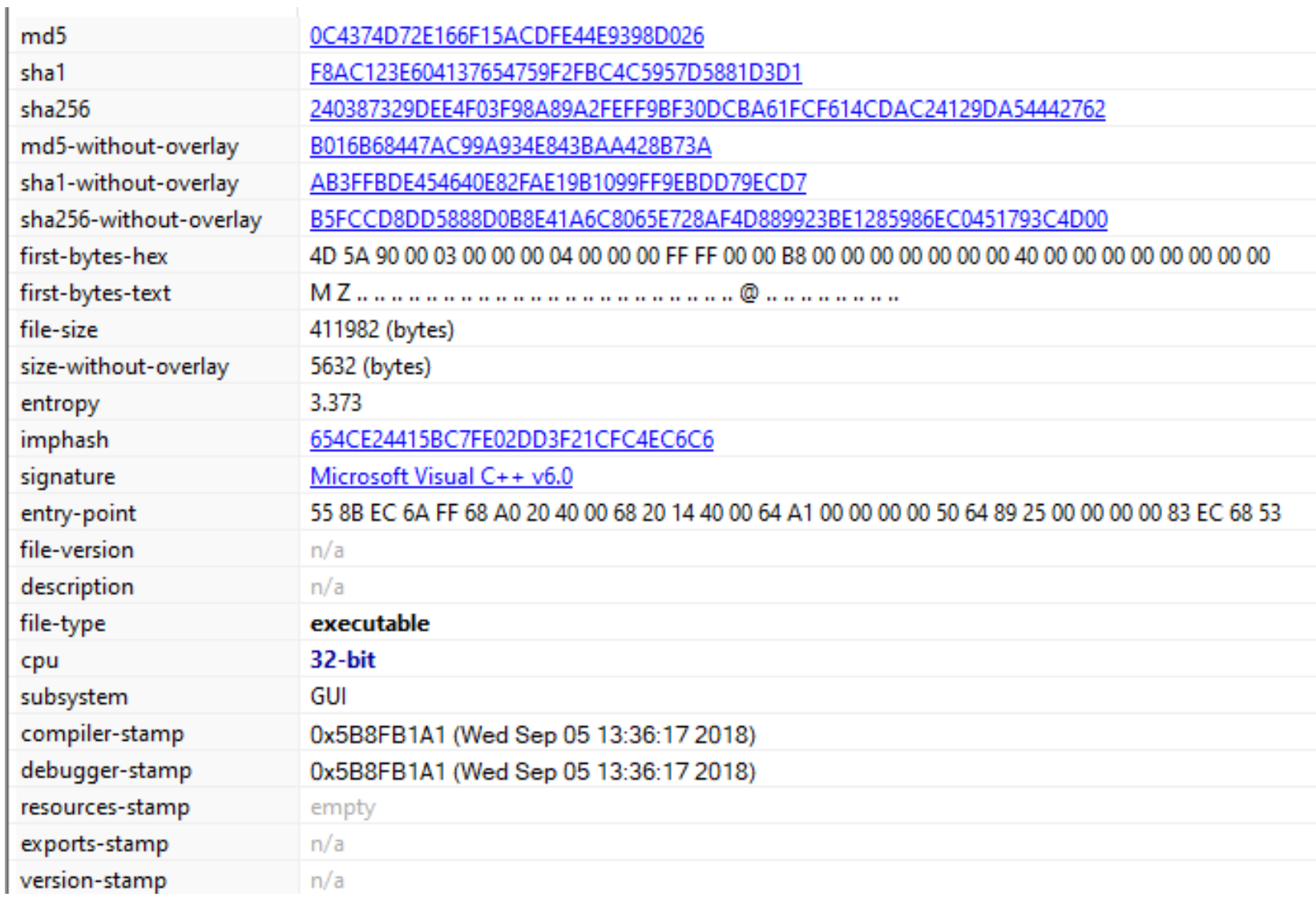

| | |
|---|---|
| md5 | 0C4374D72E166F15ACDFE44E9398D026 |
| sha1 | F8AC123E604137654759F2FBC4C5957D5881D3D1 |
| sha256 | 240387329DEE4F03F98A89A2FEFF9BF30DCBA61FCF614CDAC24129DA54442762 |
| md5-without-overlay | B016B68447AC99A934E843BAA428B73A |
| sha1-without-overlay | AB3FFBDE454640E82FAE19B1099FF9EBDD79ECD7 |
| sha256-without-overlay | B5FCCD8DD5888D0B8E41A6C8065E728AF4D889923BE1285986EC0451793C4D00 |
| first-bytes-hex | 4D 5A 90 00 03 00 00 00 04 00 00 00 FF FF 00 00 B8 00 00 00 00 00 00 00 40 00 00 00 00 00 00 00 00 |
| first-bytes-text | M Z .. .. .. .. .. .. .. .. .. .. .. .. .. .. .. .. .. .. .. .. .. .. .. .. .. @ .. .. .. .. .. .. .. .. |
| file-size | 411982 (bytes) |
| size-without-overlay | 5632 (bytes) |
| entropy | 3.373 |
| imphash | 654CE24415BC7FE02DD3F21CFC4EC6C6 |
| signature | Microsoft Visual C++ v6.0 |
| entry-point | 55 8B EC 6A FF 68 A0 20 40 00 68 20 14 40 00 64 A1 00 00 00 00 50 64 89 25 00 00 00 00 83 EC 68 53 |
| file-version | n/a |
| description | n/a |
| file-type | **executable** |
| cpu | **32-bit** |
| subsystem | GUI |
| compiler-stamp | 0x5B8FB1A1 (Wed Sep 05 13:36:17 2018) |
| debugger-stamp | 0x5B8FB1A1 (Wed Sep 05 13:36:17 2018) |
| resources-stamp | empty |
| exports-stamp | n/a |
| version-stamp | n/a |

Figure 5. Initial sample info using PEStudio

- Its imports were viewed. Hints of having a functionality related to enumeration of active services, retarding process execution, gain control of processes and more, were found.

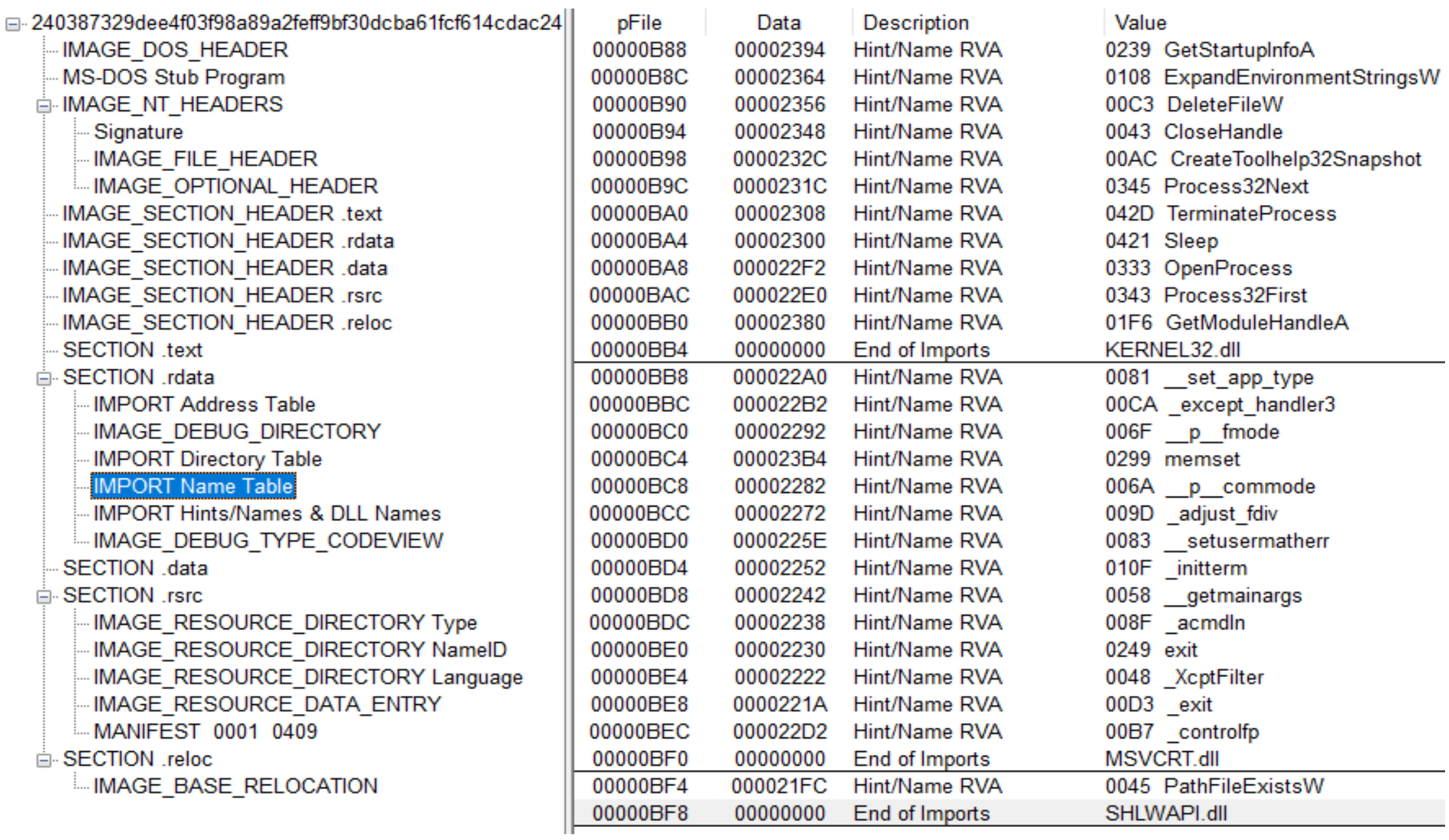

| pFile | Data | Description | Value |
|---|---|---|---|
| 00000B88 | 00002394 | Hint/Name RVA | 0239 GetStartupInfoA |
| 00000B8C | 00002364 | Hint/Name RVA | 0108 ExpandEnvironmentStringsW |
| 00000B90 | 00002356 | Hint/Name RVA | 00C3 DeleteFileW |
| 00000B94 | 00002348 | Hint/Name RVA | 0043 CloseHandle |
| 00000B98 | 0000232C | Hint/Name RVA | 00AC CreateToolhelp32Snapshot |
| 00000B9C | 0000231C | Hint/Name RVA | 0345 Process32Next |
| 00000BA0 | 00002308 | Hint/Name RVA | 042D TerminateProcess |
| 00000BA4 | 00002300 | Hint/Name RVA | 0421 Sleep |
| 00000BA8 | 000022F2 | Hint/Name RVA | 0333 OpenProcess |
| 00000BAC | 000022E0 | Hint/Name RVA | 0343 Process32First |
| 00000BB0 | 00002380 | Hint/Name RVA | 01F6 GetModuleHandleA |
| 00000BB4 | 00000000 | End of Imports | KERNEL32.dll |
| 00000BB8 | 000022A0 | Hint/Name RVA | 0081 __set_app_type |
| 00000BBC | 000022B2 | Hint/Name RVA | 00CA _except_handler3 |
| 00000BC0 | 00002292 | Hint/Name RVA | 006F __p__fmode |
| 00000BC4 | 000023B4 | Hint/Name RVA | 0299 memset |
| 00000BC8 | 00002282 | Hint/Name RVA | 006A __p__commode |
| 00000BCC | 00002272 | Hint/Name RVA | 009D _adjust_fdiv |
| 00000BD0 | 0000225E | Hint/Name RVA | 0083 __setusermatherr |
| 00000BD4 | 00002252 | Hint/Name RVA | 010F _initterm |
| 00000BD8 | 00002242 | Hint/Name RVA | 0058 __getmainargs |
| 00000BDC | 00002238 | Hint/Name RVA | 008F _acmdln |
| 00000BE0 | 00002230 | Hint/Name RVA | 0249 exit |
| 00000BE4 | 00002222 | Hint/Name RVA | 0048 _XcptFilter |
| 00000BE8 | 0000221A | Hint/Name RVA | 00D3 _exit |
| 00000BEC | 000022D2 | Hint/Name RVA | 00B7 _controlfp |
| 00000BF0 | 00000000 | End of Imports | MSVCRT.dll |
| 00000BF4 | 000021FC | Hint/Name RVA | 0045 PathFileExistsW |
| 00000BF8 | 00000000 | End of Imports | SHLWAPI.dll |

Figure 6. Sample's import table as seen through PEView

- Its strings were viewed. Other than a padding space (which can be used for lowering entropy) nothing interesting was found.

```
00000b00   0000000a   A   wincfg.exe
00000b0c   0000000c   A   wincfg32.exe
00000b1c   0000000a   A   winupd.exe
00000b28   0000000c   A   winupd32.exe
00000bfe   0000000f   A   PathFileExistsW
00000c0e   0000000b   A   SHLWAPI.dll
00000c1c   00000005   A   _exit
00000c24   0000000b   A   _XcptFilter
00000c3a   00000007   A   _acmdln
00000c44   0000000d   A   __getmainargs
00000c54   00000009   A   _initterm
00000c60   00000010   A   __setusermatherr
00000c74   0000000c   A   _adjust_fdiv
00000c84   0000000c   A   __p__commode
00000c94   0000000a   A   __p__fmode
00000ca2   0000000e   A   __set_app_type
00000cb4   00000010   A   _except_handler3
00000cc6   0000000a   A   MSVCRT.dll
00000cd4   0000000a   A   _controlfp
00000ce2   0000000e   A   Process32First
00000cf4   0000000b   A   OpenProcess
00000d02   00000005   A   Sleep
00000d0a   00000010   A   TerminateProcess
00000d1e   0000000d   A   Process32Next
00000d2e   00000018   A   CreateToolhelp32Snapshot
00000d4a   0000000b   A   CloseHandle
00000d58   0000000b   A   DeleteFileW
00000d66   00000019   A   ExpandEnvironmentStringsW
00000d82   00000010   A   GetModuleHandleA
00000d96   0000000f   A   GetStartupInfoA
00000da6   0000000c   A   KERNEL32.dll
00000db6   00000006   A   memset
00000dd8   00000032   A   C:\Users\x\Desktop\Home\Code\Mfix\Release\Mfix.pdb
00001258   00000049   A   <assembly xmlns="urn:schemas-microsoft-com:asm.v1" manifestVersion="1.0">
000012a3   00000036   A     <trustInfo xmlns="urn:schemas-microsoft-com:asm.v3">
000012db   0000000e   A       <security>
000012eb   0000001b   A         <requestedPrivileges>
00001308   0000005e   A           <requestedExecutionLevel level="asInvoker" uiAccess="false"></requestedExecutionLevel>
00001368   0000001c   A         </requestedPrivileges>
00001386   0000000f   A       </security>
00001397   0000000e   A     </trustInfo>
000013a7   00000059   A   </assembly>PAPADDINGXXPADDINGPADDINGXXPADDINGPADDINGXXPADDINGPADDINGXXPADDINGPADDINGXXPAD
0000140d   00000005   A   1>1b1
0000141f   00000015   A   2▯222272J2O2b2g2z2▯2
0000144d   00000015   A   3▯33)3.383E3W3\3a3n3
0000146d   0000000d   A   4"4(4.444:4@4
000015f7   00000056   A   "/108/99/111/113/29/41/56/31/39/55/18/16/10/54/58/44/47/34/35/63/102/14/65/109/103/" ;
```

Figure 7. Sample's strings using Detect It Easy

- All of the above findings were verified by using other relevant analysis tools.

In comparison, the second phase of MARE is the part of Isolation & Extraction. As with its Detection phase, it seems that MARE loses relevance. As already stated, MARE was more generic in terms of how it perceived, at the time of its development, MA. This is not the actual case now since there is no need for the analyst to keep in mind how the Isolation & Extraction will be conducted since it is not his/her part to do so. It is something that precedes and thus it loses relevance. Analysts are focused on the analysis and dissection of the specimen handed over to them.

### 5.1.3. Code Analysis

Next comes the Code Analysis phase. In this phase analysts perform the reverse engineering of the specimen to determine its inner workings and functionality. Code Analysis involves both static and dynamic analysis of the specimen code. This is an extremely elaborate and time-consuming phase. What we did was:

- To disassemble the sample to analyze it statically. We used IDA Free for this part. The static code analysis gave us some useful insight. Specifically, there were indications that specimen developers tried to create a "buffer" zone of actions prior to the execution of the malware. This was derived from the fact that the specimen did not load starting from the entry point. We then observed that the specimen tried to map the present processes of the system, trying to gain control and retarding execution of them (this amplifies the assumptions we made during the Classification phase). The most significant evidence found was the relation of the specimen functionality with call error functions of the system and the control of error messages [31] [32].

```
.text:0040120C 224                 push    eax             ; lpFileName
.text:0040120D 228                 call    ds:DeleteFileW
.text:00401213
.text:00401213     loc_401213:                             ; CODE XREF: sub_401150+B4↑j
.text:00401213 224                 push    3E8h            ; dwMilliseconds
.text:00401218 228                 call    ds:Sleep
.text:0040121E 224                 push    offset aWinupdsrvcExe ; "winupdsrvc.exe"
.text:00401223 228                 call    sub_401000
.text:00401228 228                 add     esp, 4
.text:0040122B 224                 push    3E8h            ; dwMilliseconds
.text:00401230 228                 call    ds:Sleep
.text:00401236 224                 push    offset aWincfgExe ; "wincfg.exe"
.text:0040123B 228                 call    sub_401000
.text:00401240 228                 add     esp, 4
.text:00401243 224                 push    3E8h            ; dwMilliseconds
.text:00401248 228                 call    ds:Sleep
.text:0040124E 224                 push    offset aWincfg32Exe ; "wincfg32.exe"
.text:00401253 228                 call    sub_401000
.text:00401258 228                 add     esp, 4
.text:0040125B 224                 push    3E8h            ; dwMilliseconds
.text:00401260 228                 call    ds:Sleep
.text:00401266 224                 push    offset aWinupdExe ; "winupd.exe"
.text:0040126B 228                 call    sub_401000
.text:00401270 228                 add     esp, 4
.text:00401273 224                 push    3E8h            ; dwMilliseconds
.text:00401278 228                 call    ds:Sleep
.text:0040127E 224                 push    offset aWinupd32Exe ; "winupd32.exe"
.text:00401283 228                 call    sub_401000
.text:00401288 228                 add     esp, 4
.text:0040128B 224                 push    3E8h            ; dwMilliseconds
.text:00401290 228                 call    ds:Sleep
.text:00401296 224                 xor     eax, eax
```

Figure 8. Retarding system processes (IDA Free)

```
.text:004012A5 008                 push    offset stru_4020A0
.text:004012AA 00C                 push    offset _except_handler3
.text:004012AF 010                 mov     eax, large fs:0
.text:004012B5 010                 push    eax
.text:004012B6 014                 mov     large fs:0, esp
.text:004012BD 014                 sub     esp, 68h
.text:004012C0 07C                 push    ebx
.text:004012C1 080                 push    esi
.text:004012C2 084                 push    edi
.text:004012C3 088                 mov     [ebp+ms_exc.old_esp], esp
.text:004012C6 088                 xor     ebx, ebx
.text:004012C8 088                 mov     [ebp+ms_exc.registration.TryLevel], ebx
.text:004012CB 088                 push    2
.text:004012CD 08C                 call    ds:__set_app_type
.text:004012D3 08C                 pop     ecx
.text:004012D4 088                 or      dword_403030, 0FFFFFFFFh
.text:004012DB 088                 or      dword_403034, 0FFFFFFFFh
.text:004012E2 088                 call    ds:__p__fmode
.text:004012E8 088                 mov     ecx, dword_40302C
.text:004012EE 088                 mov     [eax], ecx
.text:004012F0 088                 call    ds:__p__commode
.text:004012F6 088                 mov     ecx, dword_403028
.text:004012FC 088                 mov     [eax], ecx
.text:004012FE 088                 mov     eax, ds:_adjust_fdiv
.text:00401303 088                 mov     eax, [eax]
.text:00401305 088                 mov     dword_403038, eax
.text:0040130A 088                 call    nullsub_1
.text:0040130F 088                 cmp     dword_403010, ebx
.text:00401315 088                 jnz     short loc_401323
.text:00401317 088                 push    offset sub_40141C
.text:0040131C 08C                 call    ds:__setusermatherr
.text:00401322 08C                 pop     ecx
.text:00401323
```

Figure 9. Utilization of errors and I/O (IDA Free)

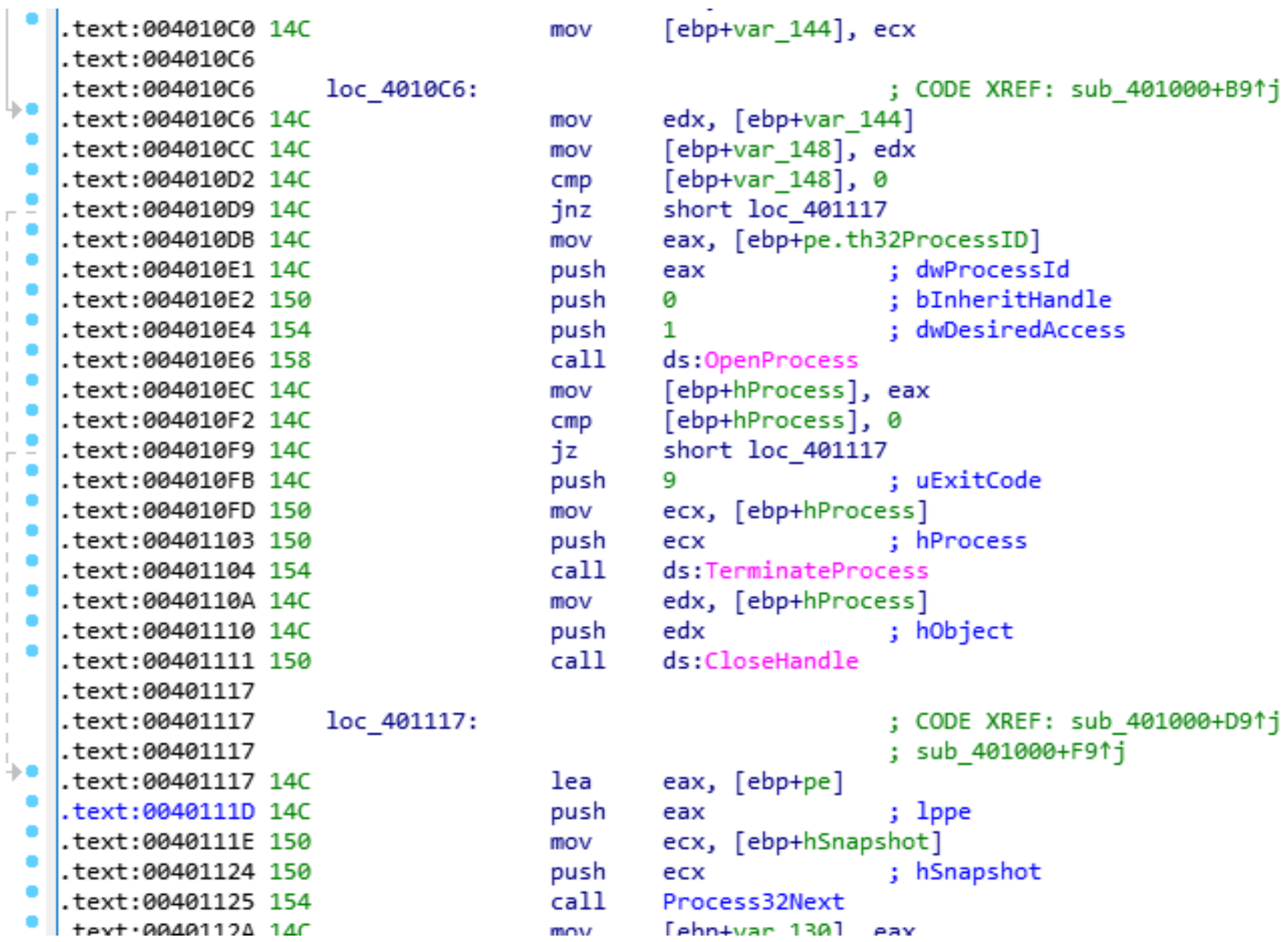

Figure 10. Manipulation of processes (IDA Free)

- To debug the sample to analyze it dynamically. We used it x32dbg for this part (we already knew from the Classification phase that the sample is a 32-bit executable). Through this process we verified the findings of the disassembling that took place before. We viewed the calling of the functions related to functionality discussed in disassembling with active handles that gave hints of image file handling (verifying assumptions on GUI nature of the specimen and that handling of system error messages) [31] [32].

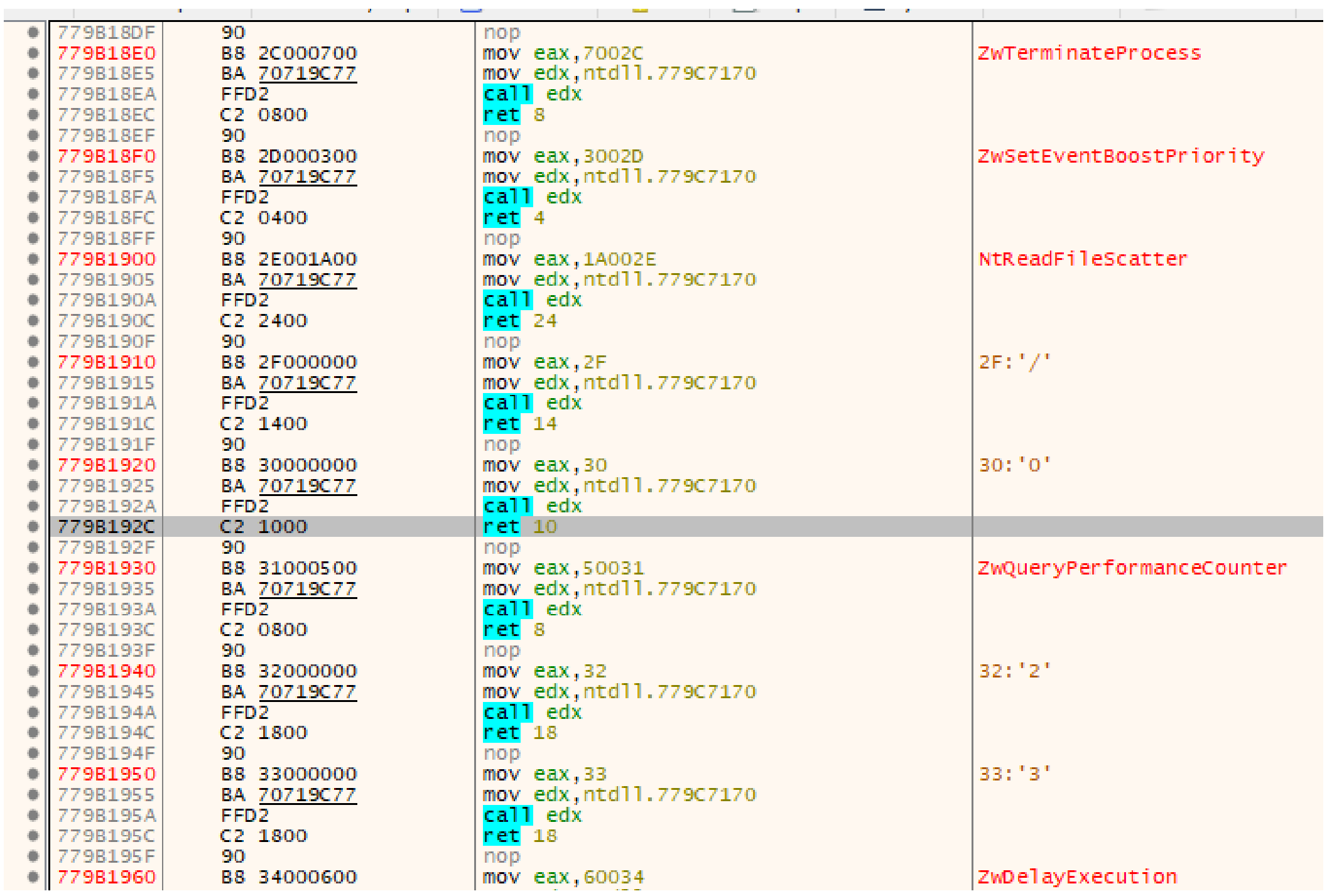

Figure 11. Calling functions related to functionality discussed in disassembling (x32dbg)1/2

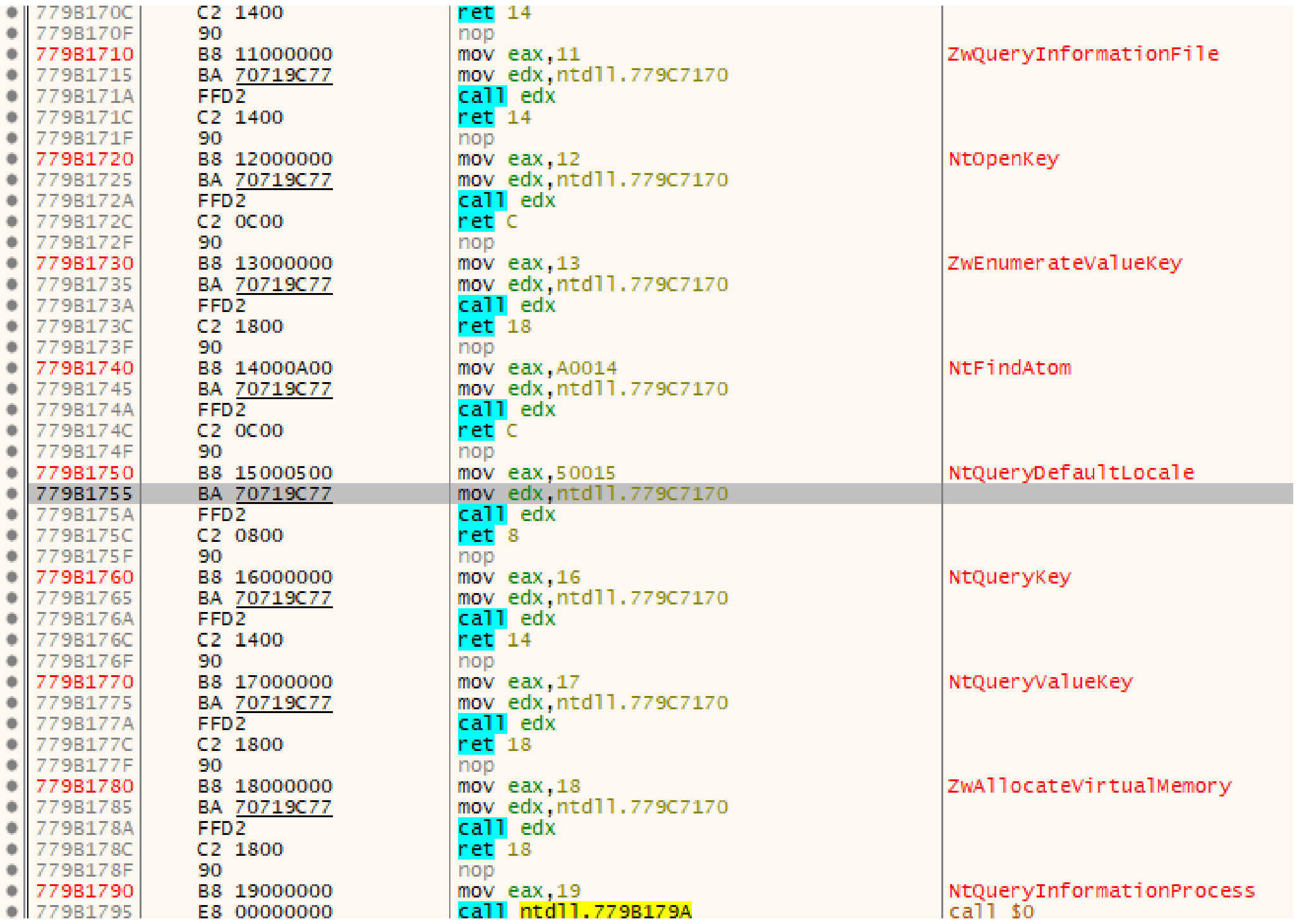

Figure 12. Calling functions related to functionality discussed in disassembling (x32dbg) 2/2

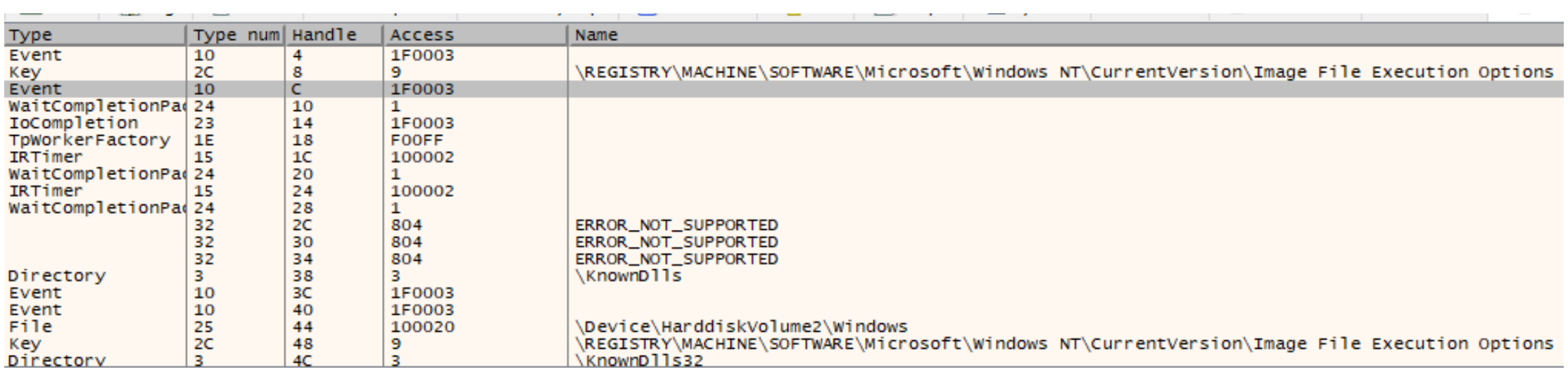

| Type | Type num | Handle | Access | Name |
|---|---|---|---|---|
| Event | 10 | 4 | 1F0003 | |
| Key | 2C | 8 | 9 | \REGISTRY\MACHINE\SOFTWARE\Microsoft\Windows NT\CurrentVersion\Image File Execution Options |
| Event | 10 | C | 1F0003 | |
| WaitCompletionPa | 24 | 10 | 1 | |
| IoCompletion | 23 | 14 | 1F0003 | |
| TpWorkerFactory | 1E | 18 | F00FF | |
| IRTimer | 15 | 1C | 100002 | |
| WaitCompletionPa | 24 | 20 | 1 | |
| IRTimer | 15 | 24 | 100002 | |
| WaitCompletionPa | 24 | 28 | 1 | |
| | 32 | 2C | 804 | ERROR_NOT_SUPPORTED |
| | 32 | 30 | 804 | ERROR_NOT_SUPPORTED |
| | 32 | 34 | 804 | ERROR_NOT_SUPPORTED |
| Directory | 3 | 38 | 3 | \KnownDlls |
| Event | 10 | 3C | 1F0003 | |
| Event | 10 | 40 | 1F0003 | |
| File | 25 | 44 | 100020 | \Device\HarddiskVolume2\Windows |
| Key | 2C | 48 | 9 | \REGISTRY\MACHINE\SOFTWARE\Microsoft\Windows NT\CurrentVersion\Image File Execution Options |
| Directory | 3 | 4C | 3 | \KnownDlls32 |

Figure 13. Active handles give hints of image file handling – indicating the tampering of GUI or graphics in error messages (x32dbg)

In comparison, the third phase of MARE is Behavioral Analysis. We can note here that both SAMA and MARE share Code Analysis and Behavioral Analysis phases. The difference is that in MARE the Behavioral Analysis phase precedes the Code Analysis one. This is another point where SAMA innovates. SAMA goes through the Code Analysis first because it takes into consideration the fact that there may be insights derived from Code Analysis that may be useful when Behavioral Analysis takes place (e.g., special comparison strings for fields found hardcoded during Code Analysis). On the other hand, there is no such case present in the precedence of Behavioral Analysis and thus we can conclude that MARE developers, while following the M.D. timeline, chose the Code Analysis to come last because of the fact that it is the most complex phase of them all.

### 5.1.4. Behavioral Analysis

The last phase of SAMA is the Behavioral Analysis phase. During this phase analysts try to view and record they changes the specimen conducts upon execution to the victim system. It is a dynamic analysis procedure and as stated it involves the execution of the Malware executable. What we did were the following:

- After the debugging phase concluded (which itself is dynamic and involves the execution of the specimen) we reverted the system to a clean state.
- We took a snapshot of the system registry using RegShot so as to have a reference to compare to after the execution of the specimen takes place.
- Then we started all the monitoring tools we had in our toolbox because we wanted to be able to observe the behavior of the specimen in great detail. We used Process Monitor to view the modifications made by processes to the system in real time, Process Explorer to observe process execution and properties, FakeNet-NG to serve virtual network services to the malware as if the system was actually online and finally, we started Wireshark to observer connections and their status in real time.
- Then we executed the specimen and we were very careful and observant to locate its execution. It is not a rare phenomenon for malware to show their execution for a really short amount of time and then disappear to let other elements setup by them continue the mission. We successfully observed its execution for a couple of seconds by the utilization of Process Explorer.

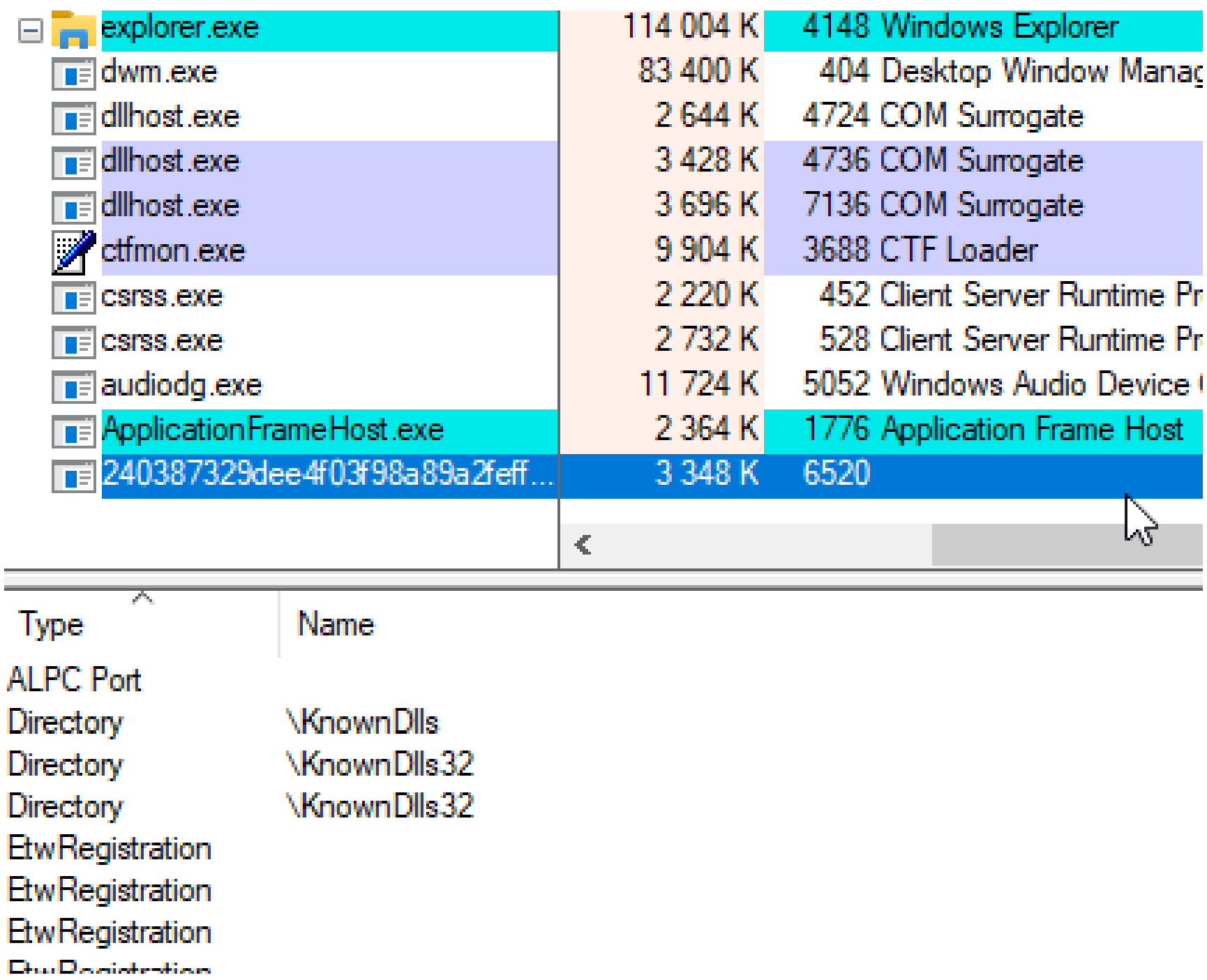

Figure 14. Malware execution as viewed through Process Explorer

- We let the specimen run for at least 20 minutes (we let it about half an hour). This is done because many specimens have defense mechanisms for avoiding detection which mechanisms delay the execution of the malware by a significant amount of time.
- Then we took another system registry snapshot to compare with the initial that we got pre-execution.

At this point it would be wise to note that SAMA suggests the integration of Memory Analysis as part of the Behavioral Analysis phase. A great amount of information can be mined from memory analysis and it is another point where SAMA shines since memory analysis is an alternative and contemporary procedure which was not streamlined when MARE was developed. For our part, due to technical difficulties, we did not proceed to memory analysis although efforts were done to obtain the relevant memory dump [30].

In comparison, the final phase of MARE is the Code Analysis & Reverse Engineering phase. As already stated, it is actually the same as the third phase of SAMA which is the Code Analysis phase. The advantages of the SAMA approach have been already stated at the previous comparisons of steps.

To conclude, what was done above (except from the introductory part of the paper), is to effectively compare both theoretically and practically the only two existent and established MA methodologies and to derive the result that SAMA is indeed an improvement of the MARE one. By using a random and modern specimen we tried to preserve the neutral and integral nature of this comparison.

## 6. CONCLUSIONS

From the dawn of MA, there were not any generic and strictly structured procedures regarding the process of it. Until now there was only one defined model and that was the MARE (Malware Analysis and Reverse Engineering). Although quite complete and effective on its own part, MARE was designed in a time where a variety of present threats were nonexistent. Recently, a new model named SAMA (Systematic Approach to Malware Analysis) emerged to the surface. SAMA analyzes MARE and reshuffles its parts while optimizing the whole process in order to gain competence in combating the contemporary threats. Ultimately, we try to understand whether SAMA is a better fit for the malware analyst of today. It seems that this is indeed the case. SAMA takes into consideration the fact that in highly sophisticated cases of threats parts of the functionality can only be observed under extreme circumstances and thus it is better to dissect the threat prior to its execution. This is why Code Analysis precedes the Behavioral Analysis. Also, by the time MARE was written the most notorious threat were rootkits and thus big part of it was written based on this fact. Rootkits still consist a formidable foe but they do not carry the same dynamic as they did previously because the malware development has shifted its focus towards other forms of threats. This is something that SAMA remedies. Finally, another fact that differentiates SAMA from MARE is that SAMA introduces into the structure of methodology alternative and complementary techniques like memory analysis. More specifically, SAMA introduces memory analysis as an integral part of it which is embedded as a part of behavioral analysis.

While tremendously useful and important, MARE seems like it started showing its age. SAMA continues the evolution of MA methodology from the point that MARE left it. SAMA adapts the MA methodology to the current threat landscape and makes it competent to cover the needs of today's analysts. It also removes all the unnecessary and redundant steps from M.D. Timeline which were included in MARE but do not constitute part of the MA itself (and as a result can be seen as irrelevant with the work of the analysts).Thus, it can be said that SAMA can be considered as an optimization of MARE. As such, SAMA achieves its purpose, and it is indeed a modern methodology for the modern MA.

As a closure we can argue that SAMA is still quite fresh and it must prove itself against highly esteemed opponents like highly sophisticated ransomware. Also, given the fact that MA evolves and extends using ML methods it is quite interesting (and also food for thought) to see whether SAMA has the required modularity to embed ML techniques in its proposed model of MA. Topics of IMG-based or entropy-based analysis, especially while deploying Deep Learning techniques has an enormous interest and it can be seen as a great path to pursue in terms of research.